\documentclass[times]{nlaauth}

\renewcommand{\DOI}{nla.2220}

\usepackage[utf8]{inputenc}

\usepackage[absolute,overlay]{textpos}
\usepackage{setspace}
\usepackage{amsmath}
\usepackage{hyperref}
\usepackage{tikz}
\usepackage{amssymb}
\usepackage{algpseudocode,algorithm,algorithmicx}

\algrenewcommand\algorithmicrequire{\textbf{Precondition:}}

\begin{document}

\runningheads{M.\,Schreiber and R.\,Loft}{Parallel Time-Integrator for Linearized SWE on Rotating Sphere}

\title{A Parallel Time-Integrator for Solving the Linearized Shallow Water Equations on the Rotating Sphere}

\author{Martin Schreiber\corrauth\affil{1}\affil{2},
Richard Loft\affil{3}}

\address{
	\affilnum{1}
	College of Engineering, Mathematics and Physical Sciences, University of Exeter, Exeter, UK
	\break
	\affilnum{2}
	Chair of Computer Architecture and Parallel Systems, Technical University of Munich, Germany
	\break
	\affilnum{3}CISL/NCAR, Boulder, Colorado, USA}
	\corraddr{Technical University of Munich, Boltzmannstrasse 3, 85748 Garching, Germany. E-Mail: martin.schreiber@tum.de}

\begin{abstract}
With the stagnation of processor core performance, further reductions in the time-to-solution for geophysical fluid problems are becoming increasingly difficult with standard time integrators. Parallel-in-time exposes and exploits additional parallelism in the time dimension which is inherently sequential in traditional methods.

The rational approximation of exponential integrators (REXI) method allows taking arbitrarily long time steps based on a sum over a
number of decoupled complex PDEs that can be solved independently massively parallel.
Hence REXI is assumed to be well suited for modern massively parallel super computers which are currently trending.

To date the study and development of the REXI approach has been limited to linearized problems on the periodic 2D plane.
This work extends the REXI time stepping method to the linear shallow-water equations (SWE) on the rotating sphere, thus moving the method one step closer to solving fully nonlinear fluid problems of geophysical interest on the sphere.
The rotating sphere poses particular challenges for finding an efficient solver due to the zonal dependence of the Coriolis term.

Here we present an efficient REXI solver based on spherical harmonics, showing the results of: a geostrophic balance test; a comparison with alternative time stepping methods; an analysis of dispersion relations, indicating superior properties of REXI; and finally a performance comparison on Cheyenne supercomputer.
Our results indicate that REXI is not only able to take larger time steps, but that REXI can also be used to gain higher accuracy and significantly reduced time-to-solution compared to currently existing time stepping methods.
\end{abstract}

\keywords{
parallel-in-time,
massively parallel,
rational approximation of exponential integrator,
shallow-water equations on the rotating sphere,
high-performance computing,
spherical harmonics
}

\maketitle

\begin{textblock*}{20cm}(3cm,2.7cm) 
	\begin{footnotesize}
		\noindent
			This is a self-archived version of the paper which has been published in final form at \url{https://onlinelibrary.wiley.com/doi/abs/10.1002/nla.2220}\\
			This article may be used for non-commercial purposes in accordance with Wiley Terms and Conditions for Use of Self-Archived Versions
	\end{footnotesize}
\end{textblock*}

\section{Introduction}

The current trend in computer architectures is towards increasing levels of parallelism while maintaining constant or even reducing individual core performance.
These developments, in place now for over a decade, pose new challenges for the development of faster solvers for nonlinear partial differential equation problems.
Such problems are found in the solvers, or dynamical cores, used in modeling the geophysical fluids in the areas of climate and weather, etc. Accelerating such problems in the post Dennard scaling\cite{dennard1974design} 
\footnote{Constraints of close-to-molecular scale from physics restrict further shrinking of components}
era is a grand challenge of the highest societal importance, and is the focus of this paper.

Overcoming the time step restrictions of time integration methods while maintaining sufficient accuracy require more sophisticated time-integration approaches, which generally focus on taking bigger time steps. Parallel-in-time methods are one class of such methods (see \cite{Gander15} for an overview).
A scalable parallel-in-time method called the rational approximation of an exponential integrator (T-REXI\footnote{Here we would like to prefix the term REXI with ``T-'' to emphasize that the REXI method of \textit{T}erry Haut et al.\, is used here. Alternative REXI-like formulations can be e.g. also based on \,Laplace transformations.}) was recently developed and explored for simulations of linear operators on the bi-periodic plane \cite{Haut2015,schreiber2016beyond}.
In this work, we will investigate the application of T-REXI to the linear shallow-water equations (SWE) on the rotating sphere. The nonlinear SWEs are an important test case in the development of dynamical cores in the atmospheric sciences: thus this work brings the T-REXI method significantly closer to modeling of planetary-scale geophysical fluids problems.

\subsection{Related work}

An alternative to the T-REXI approach to solving the linear SWE on the rotating sphere is computing the solution for the linear operator directly with an eigenvalue/vector factorization, resulting in a set of independent ODEs (see e.g.\,\cite{hochbruck2010exponential}), which can then be trivially solved with a direct time integration of the form $\hat{u}(t) = exp(\lambda t) \hat{u}(0)$ with $\hat{u}(t)$ the solution at time $t$ after transformation to eigenvector space and $\lambda$ the corresponding eigenvalue.
An eigenfactorization based on using Spherical Harmonics (SH) for the basis leads to the \textit{Hough functions} (see e.g.\,\cite{kasahara1977numerical,wang2016computation}) which are eigenfunctions of the SWE on the rotating sphere.
However, the matrices of the discretized eigenfunctions are dense and therefore have storage requirements which are quadratic in the number of degrees of freedom, i.e. the total number of SH basis functions kept in the expansion.
Running computations with the Hough functions is indeed feasible for very low resolutions, but inefficient for high resolution models due to these quadratic storage and related computational requirements.

One of the main focus of the present work is to overcome time step restrictions in atmospheric simulations.
Besides the CFL restrictions mentioned earlier,  the mathematical properties of the SWEs lead to further restrictions: specifically one finds that the gravitation-related parts of the equations are the fastest evolving and thus the most restrictive ones, requiring about six times smaller time steps compared to the meteorological modes (e.g.\,modes which are related to Rossby waves, see \cite{robert1969integration}).
Broadly speaking, two different approaches have been identified to overcome the time step restrictions of the SWEs on the rotating sphere.
The first method tackles the gravity wave issue through a semi-implicit formulation studied by Robert \cite{robert1969integration} for global spectral methods, see also \cite{hack1992description} for a discussion. Here, the stiff (linear) terms that limit the time step through the fastest gravity wave speeds are treated with an implicit time stepping approach which effectively slows down or damps these wave modes, depending on the type of implicit time integration.
While the semi-implicit approach allows significantly larger time steps, it also introduces errors which significantly increase with the time step size.
The second method addresses the CFL constraint and is based on a Lagrangian formulation of the equations.
This formulation overcomes the Eulerian CFL time step restrictions through the advection of Lagrangian fluid parcels along flow trajectories, resulting in larger time steps\cite{ritchie1988application}.
Combinations of both above mentioned methods are frequently used in atmospheric simulations\cite{wood2014inherently,Barros1995}.
Like semi-implicit methods, Lagrangian methods induce errors which grow with the time step size and avoiding such errors is obviously very desirable.

Exponential integrators are known to avoid the time step restriction for linear operators entirely, and they can be computed in various ways\cite{Moler2003}.
However, it turns out to be challenging to find a computationally efficient formulation including the Coriolis effect which allows large time steps.
An application of exponential integrators in the context of atmospheric simulations can be e.g.\,found in \cite{garcia2014exponential}. Here, significant improvements in accuracy were observed, however performance issues under realistic test conditions were found to be a severe issue.

Parallel-in-time (PinT) methods gained increasing interest over the last 50 years and we like to refer to \cite{Gander15} for an overview.
The new T-REXI time stepping method \cite{Haut2015} studied here can be interpreted as both an exponential integrator for oscillatory systems, and as a PinT algorithm that exploits additional degrees of parallelism.
T-REXI approximates the oscillatory function $\exp(ix)$ by a sum over rational functions
$\exp(ix) \approx \sum_i{\beta_i \left( ix + \alpha_i \right)^{-1}}$
with
$\alpha_i,\beta_i \in \mathbb{C}$.
This formulation shares its algebraic structure with semi-groups\cite{hochbruck2010exponential} and Laplace transformations\cite{clancy2011laplace}.
Since T-REXI and Laplace transformations not only share the way how they are applied as a solver, but also partly share the way how the coefficients of the rational approximation are derived, we go slightly into more detail here.
First of all, both methods are based on a representation of the function to be approximated (in the present work $\exp(ix)$) in a spectral basis.
For the T-REXI and Laplace method, this is based on the Fourier and Laplace transformation, respectively.
To compute the $\alpha_i$ and $\beta_i$ coefficients for the rational approximation, T-REXI directly operates in Fourier space and uses a Fourier representation of a Gaussian basis function as a proxy to approximate $\exp(i x)$ in Fourier space, followed by merging it with a rational approximation of the Gaussian basis function.
In contrast, the Laplace transformation as it was used in e.g.\,\cite{clancy2011laplace} uses the Cauchy contour integral method.
Both methods lead to a rational approximation, hence share their way of how time stepping for a PDE is performed.

\subsection{Contribution and overview}

In this work, we apply the T-REXI method \cite{Haut2015} to the linearized SWE on the rotating sphere (Sec.\,\ref{sec:shallow_water_equations}).
Sec.\,\ref{sec:spherical_harmonics} provides a description of the SH used in this work.
This is followed by a brief introduction to T-REXI in Sec.\,\ref{sec:introduction_to_rexi} which is followed by the SH-based solver for each term in T-REXI in Sec.\,\ref{sec:rexi_for_swe_with_sh}.
The efficiency of the T-REXI method is evaluated in numerical as well as performance aspects in Sec.\,\ref{sec:numerical_tests_and_performance}:
We study T-REXI with a standard steady state test case used for the development of dynamical cores\cite{williamson1992standard}.
This is followed by a test of wave propagation over a very long time scale, which focuses on gaining a deeper understanding of wave dispersion errors in physical space.
The competitiveness to other time stepping methods is shown in the context of dispersion relations via a numerical eigenvalue analysis.
Finally, computational performance comparisons on Cheyenne supercomputer are made between T-REXI and other commonly-used time stepping methods.

\section{Shallow-water equations}
\label{sec:shallow_water_equations}

The inviscid shallow-water equations (SWE) can be described in a very general way by
$
	\frac{\partial}{\partial t} \textbf{U} = \mathcal{L}(\textbf{U}) + \mathcal{N}(\textbf{U})
$
where $\textbf{U}$ is a 3-vector containing the prognostic state variables, $\mathcal{L}(\textbf{U})$ is the linear operator, and $\mathcal{N}(\textbf{U})$ the non-linear operator.
While we recognize the critical importance of the non-linear terms in faithfully reproducing realistic atmospheric dynamics, our intent here is to extend the T-REXI work on the linear SWE on the bi-periodic plane to the rotating sphere, and determine if the approach can work competitively on the rotating sphere.

Special attention has to be paid to the choice of the formulation of the shallow-water equations.
We chose the vorticity-divergence formulation (see \cite{temperton1991scalar,hack1992description})
in which the prognostic variables are given by the geopotential $\Phi$, the vorticity $\zeta$ and divergence $\delta$
\begin{eqnarray}
	\left[\begin{array}{c}
	\frac{\partial\zeta}{\partial t}\\
	\frac{\partial\delta}{\partial t}\\
	\frac{\partial\Phi}{\partial t}
	\end{array}\right]=\left[\begin{array}{c}
	-f\delta-\textbf{V}.(\nabla f)\\
	f\zeta+\mathbf{k}.(\nabla f)\times\textbf{V}-\nabla^{2}\Phi\\
	-\overline{\Phi}\delta
	\end{array}\right]
\end{eqnarray}
with the average geopotential $\overline{\Phi}$, $\mathbf{k}$ a vector orthogonal to the surface, and the latitudinal-varying Coriolis effect is included with the $f=2\Omega\sin\phi$ term.
The longitude-latitude coordinates are denoted with angles $\lambda$ and $\phi$, respectively.
$\textbf{V} = (u, v)$ stores the velocities and is used as a diagnostic variable which can be computed based on the stream function $\psi$ and velocity potential $\chi$ via
$	\textbf{V}  =  \mathbf{k}\times\nabla\psi+\nabla\chi$.
The stream function and velocity potential can be computed by inverting the Laplace operator in the equations 
$	\zeta	=\nabla^{2}\psi~~~~\text{and}~~~~\delta	=\nabla^{2}\chi$.

\section{Spherical Harmonics}
\label{sec:spherical_harmonics}

Expansions in terms of the SH functions have long been used in atmospheric dynamics solvers, particularly to invert, algebraically, similar Helmholtz operators in the context of the semi-implicit methods, e.g.\,CAM3\cite{collins2004description}, ECMWF \cite{Barros1995,white2000ifs}. This past success, along with the prospect of an exact inversion of each term of the rational approximation of the exponential integrator, motivated our choice of SH for this formulation. 
We give a very brief introduction to SH here 
and to form the basis of a comprehensive description of the development of the T-REXI time stepping solver in Sec.\,\ref{sec:rexi_for_swe_with_sh}.

The transformation of a function $\xi(\lambda,\mu)$  from \emph{spectral} to \emph{physical space} with the Gaussian latitude $\mu = sin(\phi)$ is given by the approximate series expansion
\begin{eqnarray*}
	\xi(\lambda,\mu)=\sum_{r=-R}^{R}\sum_{s=|r|}^{S(r)}\xi_{s}^{r}P_{s}^{r}(\mu)e^{i r\lambda}
\end{eqnarray*}
with $\xi_{s}^{r}$ the spectral coefficient of the $s$-th longitudinal mode and $r$-th latitudinal mode.
In practice with a $T$-truncated SH we use a triangular truncation $(S(r)=R=T-1)$ and $s \geq r$, although more general truncations are possible.
Data from \emph{physical space} can be transformed to \emph{spectral space} with
\begin{eqnarray*}
	\xi_{s}^{r}=\int_{-1}^{+1}\frac{1}{2\pi}\int_{0}^{2\pi}\xi(\lambda,\mu)e^{-i r\lambda}d\lambda P_{s}^{r}(\mu)d\mu.
\end{eqnarray*}
The $P_s^r(\mu)$ are the associated Legendre polynomials (ALPs) and the integral over the ALPs is discretized using Gaussian quadrature and requires that the latitudes are located at the quadrature points $\mu_{j}=\sin\phi_{j}$.
The effective resolution of simulations is specified by the truncation of the modes which is abbreviated with ``$T[int]$''.
Here, T is followed by the placeholder $[int]$ for the highest Fourier and ALP mode.
Our formulation follows that of Robert \cite{robert1966integration}, who suggested scaling velocities in the SWE with $\cos\phi$.
More details on the SH can be found in \cite{hack1992description,rivier2002efficient}.

The properties of the SH are frequently used in global spectral methods to solve e.g.\,for the Laplace operator (see \cite{hack1992description,Barros1995,schreiber2016beyond}) and for Helmholtz problems arising as part of implicit time stepping.
These problems give rise to diagonal operators in spectral space.

\section{Rational Approximation of Exponential Integrators (REXI)}
\label{sec:introduction_to_rexi}

We now briefly introduce T-REXI's background and its algorithmic structure, in order to lay the groundwork for the numerical analysis in Sec.\,\ref{sec:numerical_tests_and_performance}. 

\subsection{ODEs}
\label{sec:rexi_odes}
T-REXI is composed of two main building blocks.
The first is a rational approximation of a Gaussian function
$
	\psi_{h}(x)
		:=(4\pi)^{-\frac{1}{2}}e^{-x^{2}/(4h^{2})}
		\approx  Re\left(\sum_{l=-K}^{K}\frac{a_{l}}{i\frac{x}{h}+(\mu+i\,l)}\right)
$
with the coefficients $\mu$ and $a_{l} \in \mathbb{C}$ provided in \cite[Table 1]{Haut2015} and $h$ related to the width of the Gaussian function.
An important numerical property is that $K=11$ is sufficient to keep the error of this approximation close to the level of numerical double precision accuracy\cite{Haut2015}.
The second block is a linear combination of Gaussian basis functions to approximate the spectrum of the oscillatory representation $\exp(i \lambda)$, $i \lambda$ being the eigenvalues of the underlying linear system along the imaginary axis. E.g.\,the real part of the exponential is approximated over the spectral area $[-hM;hM]$ with
\begin{eqnarray}
		Re(e^{i x}) &\approx&	\sum_{m=-M}^{M} Re\left( b_{m}\psi_{h}(x+mh) \right)
		\approx	\sum_{n} Re\left( \frac{\beta^{Re}_{n}}{ix+\alpha_{n}}\right)
\end{eqnarray}
with $Re$ the real part of a complex number.
Here, $\beta^{Re}_n, \alpha_n \in \mathbb{C}$ are complex values computed only once on initialization, and $M$ specifies the number of Gaussian basis functions used in the approximation.
The approximation for the real parts of the exponential is merged with the one for the imaginary values, yielding the coefficients $\alpha_n$ and $\beta_n$.
Finally, the sum of the rational approximation of the Gaussian function is merged with the sum for approximating the oscillations.

\subsection{PDEs}
\label{sec:rexi_pdes}

The utility of the rational approximation becomes clear when it is applied to a linear operator.
Given a linear PDE $\frac{\partial}{\partial t}\textbf{U} = L \textbf{U}$ where $L$ is the discretized linear operator and $\textbf{U}$ from now on the discretized current state of the simulation, we can compute solutions of the simulation at time $t_{n+1}$ with the exponential integrator formulation\cite{Moler2003}
$
	\textbf{U}_{n+1} =  Q
				\exp \left( \Delta t \Lambda \right)
				Q^{-1}
				\textbf{U}_{n}
$
where $\Lambda$ is a matrix with (imaginary) eigenvalues on the diagonal, $Q$ the matrix of eigenvectors, and
$
	\exp({\Delta t \Lambda})=diag(
		e^{\lambda_{1}\Delta t},
		e^{\lambda_{2}\Delta t},
		...,
		e^{\lambda_{n}\Delta t}
	)
$.
Approximating $\exp(\Delta t \Lambda)$ with T-REXI yields
$
	\textbf{U}_{n+1} \approx  Q \sum_{k=1}^{N} \left( \beta_{k} \left( {\Delta t \Lambda+I\alpha_{k}} \right)^{-1} \right) Q^{-1} \textbf{U}_{n}
			= 
		 \sum_{k=1}^{N} \beta_{k} \left( { L + I \alpha_{k} } \right)^{-1} \textbf{U}_{n}.
$
Each of these terms are independent from each other and can be solved in parallel \cite{Haut2015,schreiber2016beyond}.
We like to emphasize that REXI-based methods never require computing $Q$ and $\Lambda$ explicitly and that we use this only as an intermediate step.
It only requires solving for the complex-valued problem given by $(L+I\alpha_k)^{-1}$. Additionally, we assume that valid eigenvectors and -values exist with singular eigenvalues set to zero.
\subsection{REXI with shallow-water equations}

In this section we will develop the T-REXI formulation of the linear shallow-water equations on the rotating sphere and will derive an efficient solver using an SH representation. For sake of clarity, in what follows, we scale the time step size $\Delta t$ to unity.
We also use studies for a T-REXI formulation of the SWE on the sphere with a simplified system that assumes an unphysical, constant Coriolis force.
This modification of the SWE is sometimes called the f-sphere, and is used for numerical studies (see e.g.\,\cite{thuburn2009numerical}).
We use this f-sphere approximation only for the numerical eigenvalue analysis in Section \ref{sec:eigenvalue_analysis} since the eigenvalues can be computed explicitly for the f-sphere.
In contrast to the f-sphere, the SWE on the rotating sphere includes the true latitudinal varying Coriolis term and a similar approach for implicit time stepping methods was also independently developed in \cite{yessad2007semi}.

For T-REXI on the rotating sphere, we start with the linear operator given in advective formulation
\begin{eqnarray}
	\label{eqn:rexi_formulation_swe_on_sphere}
	\left[\begin{array}{ccc}
	\alpha & -\overline{\Phi}\nabla^{\lambda}. & -\overline{\Phi}\nabla^{\phi}.\\
	-\nabla^{\lambda} & \alpha & f(\phi)\\
	-\nabla^{\phi} & -f(\phi) & \alpha
	\end{array}\right] \mathbf{U}=\mathbf{U}^{0}
\end{eqnarray}
with $\mathbf{U}^0$ the initial condition, $\mathbf{U} = (\Phi, u, v)^T$ the state vector with the geopotential $\Phi = g h$, the gravity value $g$, height of the SWE $h$ and both velocity components $u$ and $v$.
The average geopotential is given by $\overline{\Phi} = g \overline{h}$ and $\alpha$ one of the poles in the REXI sum.
The superscripts $\lambda$ and $\phi$ denote the parts of the gradient and divergence operators along the longitude or latitude, respectively.
For sake of readability we write $f=f(\phi)$.
We can find an explicit formulation of the geopotential
with
\noindent
\begin{eqnarray}	
	\label{eqn:phi_depending_on_f}
	\left(\left(\alpha^{2}+f^{2}\right)-\overline{\Phi}\nabla^{2}\right)\Phi+\frac{\overline{\Phi}}{\alpha}F
	=
	\overline{\Phi}\left(\delta^{0}-\frac{f}{\alpha}\zeta^{0}\right)+\left(\alpha+\frac{f^{2}}{\alpha}\right)\Phi^{0}
\end{eqnarray}
and
$
	F=f\left(f_{\phi}v+f_{\lambda}u\right)+\alpha\left(f_{\lambda}v-f_{\phi}u\right)
$
where the subscripts denote the gradient along the longitude ($\lambda$) or latitude ($\phi$).
Unfortunately, using this equation directly does not result in an efficient solver in spectral space since the dependencies on $u$ and $v$ significantly increase the bandwidth and size of the matrix to solve for.
However, a formulation that only depends on the geopotential can be derived as
\begin{eqnarray}
	\label{eqn:rexi_geopotential_non_sph}
	\left(\left(\alpha^{2}+f^{2}\right)+\frac{\overline{\Phi}}{\alpha}F_{k}\nabla-\overline{\Phi}\nabla^{2}\right)\Phi=\overline{\Phi}\left(\delta^{0}-\frac{f}{\alpha}\zeta^{0}\right)+\left(\alpha+\frac{f^{2}}{\alpha}\right)\Phi^{0}-\frac{\overline{\Phi}}{\alpha}F_{k}\mathbf{V}^{0}
\end{eqnarray}
with
$\mathbf{V}^0 = (u^0, v^0)^T$ and
$
	F_{k} = \frac{1}{\alpha^{2}+f^{2}}\nabla^{\phi}f\left[\begin{array}{cc}
-\left(\alpha^{2}-f^{2}\right), & 2\alpha f\end{array}\right]
$
and $\nabla^\phi$ denoting the gradient component along the latitude.
Once the geopotential $\Phi$ at the new time step is computed, we can obtain the velocities based on Eq.\,\eqref{eqn:rexi_formulation_swe_on_sphere}
\begin{eqnarray}
	\label{eq:computing_velocities}
	\mathbf{V}=A^{-1}\left(\mathbf{V}^{0}+\nabla\Phi\right)
\end{eqnarray}
with
$
	A^{-1}=
		\frac{1}{\alpha^{2} + f^{2}}\left(\begin{array}{cc}
		\alpha & -f\\
		f & \alpha
		\end{array}\right).
$

This reformulation leads to a significant simplification of the original problem given in equation \eqref{eqn:rexi_formulation_swe_on_sphere}, which coupled all variables given by the geopotential $\Phi$ and velocities $\mathbf{V}$.
In equation \eqref{eqn:rexi_geopotential_non_sph} the predicted geopotential $\Phi$ depends only on the initial conditions, avoiding a coupling to the velocities.

\section{REXI for shallow-water equations with Spherical Harmonics}
\label{sec:rexi_for_swe_with_sh}

In this section, we turn our attention to using the SH to find an efficient solver for Eq.\,\eqref{eqn:rexi_geopotential_non_sph}. 

Our main goal is to transform Eq.\,\eqref{eqn:rexi_geopotential_non_sph} to a form in SH spectral space $\mathcal{B} \tilde{\Phi} = \tilde{r}$, where $\mathcal{B}$ is the matrix to be inverted in spectral space, the vector $\tilde{r}$, and $\tilde{\Phi}$ are spectral coefficients of the right hand side and geopotential respectively. 
SH recurrence identities
and orthonormality properties are used to compute the matrix elements of $\mathcal{B}$.
Unfortunately, Eq.\,\eqref{eqn:rexi_geopotential_non_sph} has the troublesome quantity $(\alpha^{2}+f^{2})^{-1}$ in the $F_{k}$ term, which is not amenable to evaluation using ALP identities.
Therefore, we multiply Eq.\,\eqref{eqn:rexi_geopotential_non_sph} through $(\alpha^{2}+f^{2})$ which gives
\begin{eqnarray}
\label{eqn:rexi_geopotential_mul_alpha2_f2}
	\nonumber
	\left(\left(\alpha^{2}+f^{2}\right)^{2}+\frac{\overline{\Phi}}{\alpha}\nabla^{\phi}f\left[\begin{array}{cc}
-\left(\alpha^{2}-f^{2}\right), & 2\alpha f\end{array}\right]\nabla-\left(\alpha^{2}+f^{2}\right)\overline{\Phi}\nabla^{2}\right)\Phi\\
 = \left(\alpha^{2}+f^{2}\right)\left(\overline{\Phi}\left(\delta^{0}-\frac{f}{\alpha}\zeta^{0}\right)+\left(\alpha+\frac{f^{2}}{\alpha}\right)\Phi^{0}+\frac{\overline{\Phi}}{\alpha}F_{k}\mathbf{V}^0\right).
\end{eqnarray}

The right hand side $\tilde{r}$ can be directly evaluated, and is partly assembled in physical space for the velocity-related components to avoid velocity-induced pole singularities in spectral space.
We split the left hand side into smaller subproblems denoted by $Z_n$ as follows
\begin{eqnarray*}
 &  & \left(\alpha^{2}+f^{2}\right)^{2}\Phi
 +\frac{\overline{\Phi}}{\alpha}\nabla^{\phi}f\left[\begin{array}{cc}
-\left(\alpha^{2}-f^{2}\right), & 2\alpha f\end{array}\right]\nabla\Phi
-\left(\alpha^{2}+f^{2}\right)\overline{\Phi}\nabla^{2}\Phi\\
 & = & \alpha^{4}Z_{1}+2\left(2\Omega\right)^{2}\alpha^{2}Z_{2}+\left(2\Omega\right)^{4}Z_{3}
 -\overline{\Phi}\alpha\left(2\Omega\right)Z_{4}+\frac{\overline{\Phi}}{\alpha}\left(2\Omega\right)^{3}Z_{5}\\
 & & +\overline{\Phi}2\left(2\Omega\right)^{2}Z_{6}
 -\overline{\Phi}\alpha^{2}Z_{7}-\overline{\Phi}\left(2\Omega\right)^{2}Z_{8}
\end{eqnarray*}
and find the coefficients which represent each of the $Z_i$ functions.
These functions can be derived by recurrence identities of SH.
We finally assemble our matrix $\mathcal{B}$ by collecting all coefficients associated with the same SH mode of the geopotential.
Once $\tilde{\Phi}$ has been calculated by solving for $\mathcal{B}$, the final step in the T-REXI solver is to compute the velocities or vorticity/divergence using Eq.\,\eqref{eq:computing_velocities} for each term (i.e. each value of $\alpha$) in the T-REXI series.
The vorticity and divergence $\zeta^0$, $\delta^0$ is first converted to its velocity components $u^0$, $v^0$ and the entire right-hand side is evaluated.
Then, the vorticity and divergence $\zeta$, $\delta$ is computed based on the new velocity components.
The T-REXI time stepping scheme for one single time step is summarized in Algorithm \ref{alg:rexi_algorithm}.

\begin{figure}
	\begin{center}
	\includegraphics[width=0.3\textwidth]{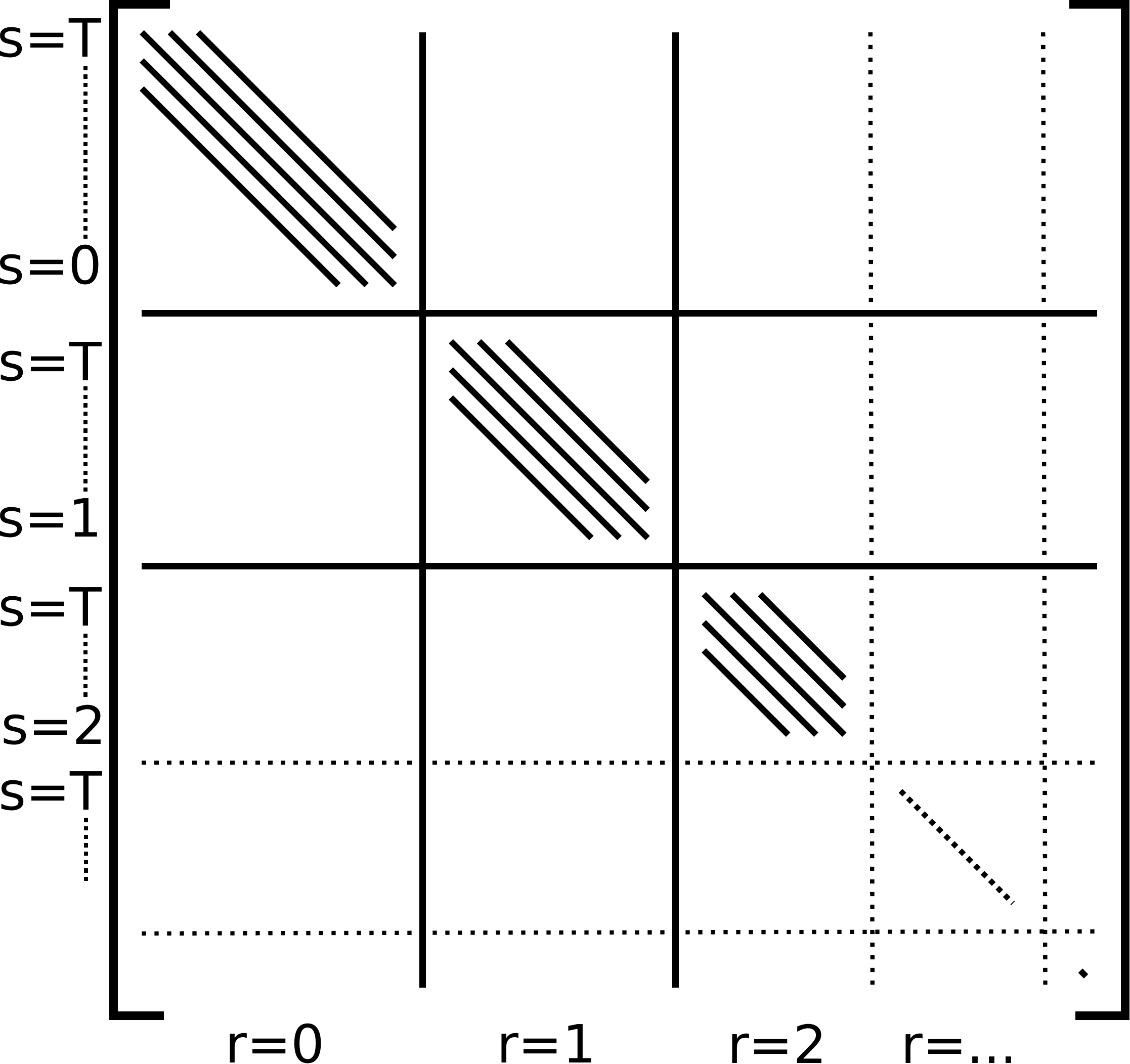}
	\end{center}
	\caption{Sketch of the matrix structure to solve for the geopotential on the rotating sphere as part of the T-REXI time stepping method.
	There are four off-diagonals with distances $\pm 2$ and $\pm 4$ to the diagonal.
	The matrix is blocked for each $m$ mode.
	Therefore, the linear system of equations in each partition can be solved independently to the others.
	$T$ denotes the truncation for the SH, see Section \ref{sec:spherical_harmonics}.
	}
	\label{fig:sketch_of_matrix_structure_phi}
\end{figure}

We close this section with a discussion on properties of the matrix $\mathcal{B}$ for which a sketch is given in Figure \ref{fig:sketch_of_matrix_structure_phi}.
This matrix has a low bandwidth $b = 5$, and can be solved using a backward/forward Gaussian elimination matrix inversion technique.
We also note that the matrix $\mathcal{B}$ is meridionally blocked, because of the pure latitude dependence of the Coriolis term.
Hence, each $m$-th zonal mode generates a system of equations which can be independently treated.
This naturally exposes an additional degree of parallelization over all $m$ modes.

\begin{algorithm}
\caption{Pseudo code for one single T-REXI time step by using the developed SH-REXI solver for vorticity/divergence formulation.}\label{alg:rexi_algorithm}
\begin{algorithmic}[1]
	\State Functions used in SHREXI solver:
	\State ~~~$S(\ldots)$: Solvers for geopotential $\Phi$ and velocities $(u,v)$
	\State ~~~$uv2vortdiv(\ldots)$: Convert velocities to vorticity/divergence
	\State ~~~$vortdiv2uv(\ldots)$: Convert vorticity/divergence to velocities
    \State
	\State $R \in (\mathbb{C}^2)^N$	\Comment{REXI coefficients for $N$ T-REXI terms}
	\State
	\Procedure{SHREXI::solve}{$\tilde{\Phi}_0, \tilde{\zeta}_0, \tilde{\delta}_0$}	\Comment{Initial conditions}
    	\State ($\tilde{\Phi}_{ret}, \tilde{\zeta}_{ret}, \tilde{\delta}_{ret}) \gets (\vec{0}, \vec{0}, \vec{0})$
		\ForAll{$(\alpha, \beta) \in R$}	\Comment{\textbf{Parallel for}}
			\State $\tilde{\Phi}_{new} \gets S^{\Phi}(\tilde{\Phi}_0, \tilde{\zeta}_0, \tilde{\delta}_0, \alpha)$	\Comment{Solve for new geopotential}
			\State $(u_0,v_0) \gets vortdiv2uv(\tilde{\zeta}_0, \tilde{\delta}_0)$	\Comment{Convert vort/div to velocities}
			\State $u_{tmp} \gets S^{u}(\tilde{\Phi}_{new}, \tilde{\Phi}_0, u_0, v_0, \alpha)$	\Comment{Solve for u velocity}
			\State $v_{tmp} \gets S^{v}(\tilde{\Phi}_{new}, \tilde{\Phi}_0, u_0, v_0, \alpha)$	\Comment{Solve for v velocity}
			\State $(\tilde{\zeta}_{new}, \tilde{\delta}_{new}) \gets uv2vortdiv(u_{tmp},v_{tmp})$	\Comment{Convert velocities to vort/div}
			\State $({\Phi}_{new}, {\zeta}_{new}, {\delta}_{new}) \gets ({\Phi}_{new}, {\zeta}_{new}, {\delta}_{new}) + Re(({\Phi}_{new}, {\zeta}_{new}, {\delta}_{new}) \beta)$ \Comment{\textbf{Parallel reduce operation}}
		\EndFor
		\State \textbf{return} $(\tilde{\Phi}_{new}, \tilde{\zeta}_{new}, \tilde{\delta}_{new})$	\Comment{Value of new time step}
	\EndProcedure
\end{algorithmic}
\end{algorithm}

\section{Numerical tests and performance results}
\label{sec:numerical_tests_and_performance}

We have conducted numerical studies to obtain a deeper understanding of T-REXI's numerical performance integrating the SWE on the rotating sphere.
If not otherwise stated, we used earth parameter values
$r=6.37122 \cdot 10^6 m$,
$\Omega=7.292 \cdot 10^{-5} s^{-1}$,
$g=9.80616 m/s^2$,
and an average height of $\overline{H} = 10000 m$.
Regarding T-REXI parameters (see Sec.\,\ref{sec:rexi_odes} and \ref{sec:rexi_pdes}), the number of Gaussian basis functions for each approximation of the real and imaginary parts of the exponential function is denoted as $M$.
Also, the studies conducted here use a value of $h=0.15$ for the width scaling parameter of the Gaussian basis functions.
For sake of reproducability, the source code is made available at \cite{sweet_repository}\footnote{repository \cite{sweet_repository}, commit from 2018-06-04, benchmarks in \texttt{benchmarks\_sphere/sph\_rexi\_linear\_paper\_*}}.
We use SHTNS \cite{schaeffer2013efficient} library for SH transformations.
For the explicit Runge-Kutta 2 method we use the coefficients $A=(0.5)$, $b=(0, 1.0)$, $c=(0.5)$ given in Butcher-tableau notation.
\subsection{Geostrophic balance}
\label{sec:geostrophic_balance}

We first investigate the T-REXI time stepping method with a test of the stationary modes generated by a geostrophic balance between the velocity and the Coriolis effect.
Considered to be a crucial test of any dynamical core\,\cite{staniforth2012horizontal,thuburn2012framework},
it is also ``test case 2'' of a standard suite of tests\cite{williamson1992standard} for the SWE and we use a linearized version of this originally non-linear benchmark.
We generate the geostrophic balance with the following initial conditions: a zero meridional velocity $v=0$, a zonal velocity
$
	u(\lambda,\phi) = u_0 \cos \phi
$
with $u_0 = 2 \pi r / 12$, and the geopotential given by
$
	\Phi(\lambda, \phi) =  g \overline{H} + u_0 r \Omega \cos^2 \phi
$.
The time derivatives generated by the geopotential balance with the Coriolis term are zero.
This benchmark is in particular interesting for studying T-REXI because it is not explicitly formulated in terms of time derivatives in contrast to e.g.\,explicit Runge-Kutta methods which solely rely on time derivatives and with the time derivatives being analytically zero for this particular benchmark.
We conducted studies over a simulation time of $1$ on the unit sphere and set $g=\overline{H}=f=1$ as well, which allows us to discriminate between round-off and time stepping errors.
We used a time step size of $\Delta t_{RK2} = 0.01$ for RK2 methods and a $10$ times larger time step size of $\Delta t_{T-REXI}=0.1$ for T-REXI.
All studies are conducted on an effective resolution of $T64$ (see Sec.\,\ref{sec:spherical_harmonics}).

\begin{figure}
	\begin{center}
	\includegraphics[width=0.9\textwidth]{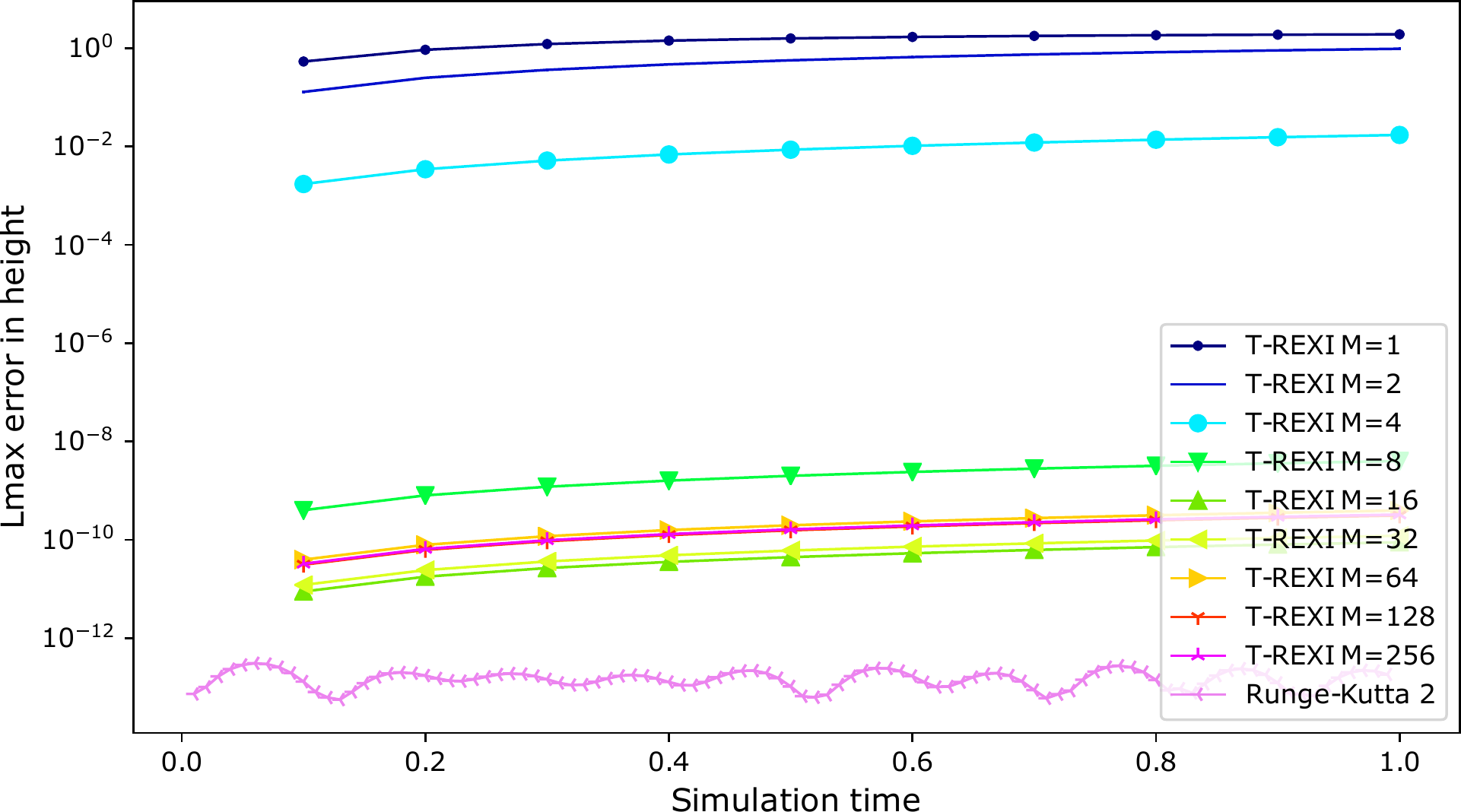}
	\end{center}
	\caption{
	\label{fig:geostrophic_benchmark_without_normalization}
		Geostrophic balance benchmark for the T-REXI coefficients which were \textit{not optimized for geostrophic balance}:
		The plot shows the maximum error of height field vs.\,simulation time.
		High accuracy for the RK2 method is apparent:
		however, the T-REXI time stepping method errors are consistently much larger, even after a single time step.
		}
\end{figure}

\begin{figure}
	\begin{center}
	\includegraphics[width=0.9\textwidth]{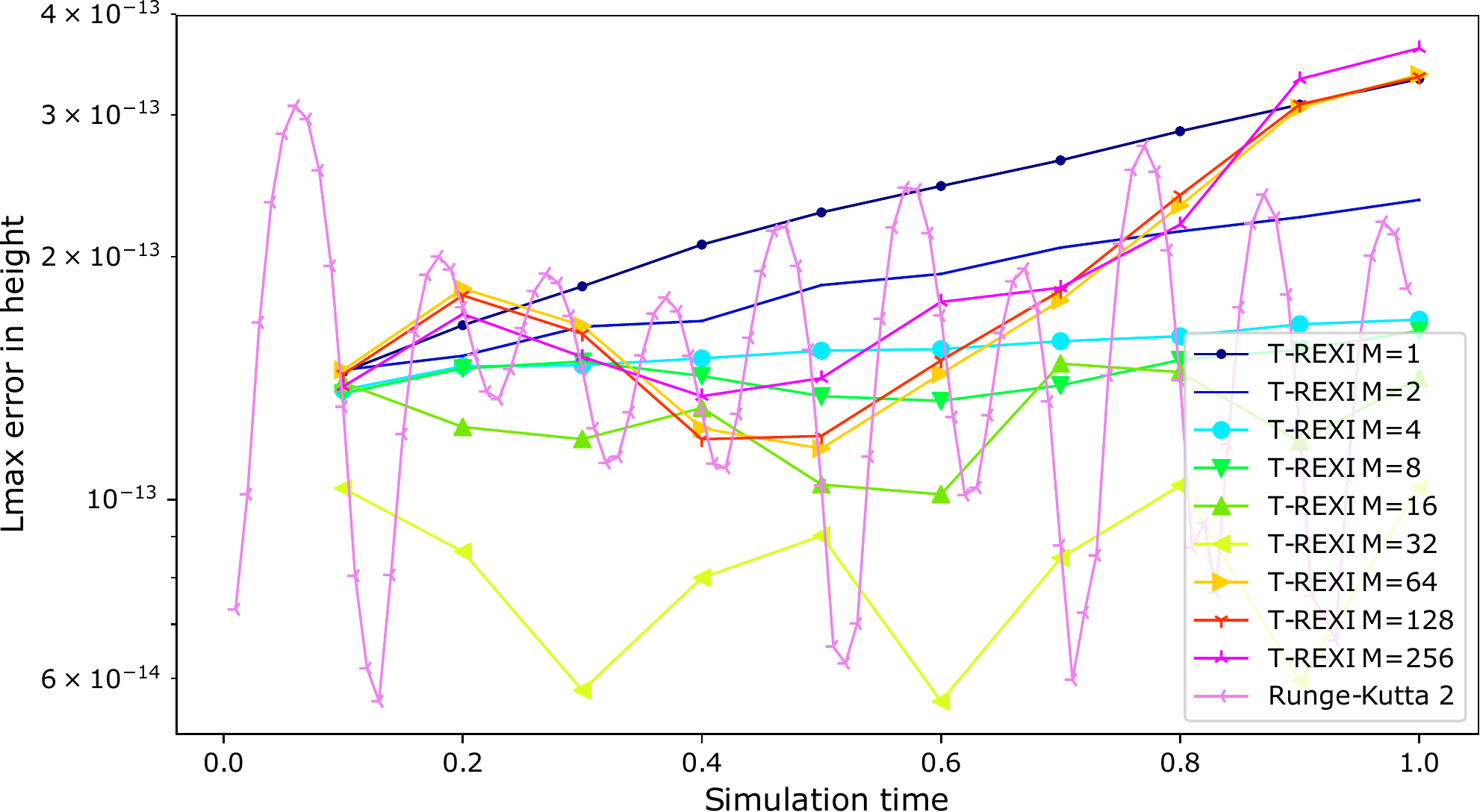}
	\end{center}
	\caption{
	\label{fig:geostrophic_benchmark_with_normalization}
		Geostrophic balance benchmark for different T-REXI coefficients $M$.
		The T-REXI poles are \textit{optimized for geostrophic balance} by renormalizing the rational approximation for accurate representation of zero eigenvalues.
		The $\ell_{\infty}$ error of the height field vs. the simulation time is shown.
		The values at the simulation time 0 are omitted.
		The T-REXI time stepping results are significantly improved up to the order of double precision accuracy.
		}
\end{figure}

The time dependent errors in the height field in the $\ell_{\infty}$ norm are given in Fig.\,\ref{fig:geostrophic_benchmark_without_normalization}. 
The T-REXI experiments were conducted with varying numbers of Gaussian basis functions $M =\{2^n | 0 \leq n \leq 8\}$.
The RK2 time stepping method shows the expected very high double precision accuracy of $O(10^{-13})$ and we like to mention here that this small error is only possible because the steady state test case is designed to provide zero time derivatives; RK2 is solely based on time derivatives, hence errors directly cancel out and significant errors with RK are visible in other benchmarks in the next sections.
T-REXI which is not based on the time derivatives displays errors which are considerably higher, even for relatively large numbers of T-REXI terms.
The source of this is a T-REXI error which was traced to a subtle issue in the coefficients used in a rational approximation  
$
	e^{ix}\approx\sum_{n}\beta_{n}(\alpha_{n}+ix)^{-1}.
$
Tests for $x=0$ (related to stationary modes which are the only ones in this benchmark) with $M=128$ T-REXI poles reveal a residual error of
\begin{eqnarray}
	1-e^{i0}\approx\sum_{n}\beta_{n}(\alpha_{n}+i0)^{-1}=1-\sum_{n}\frac{\beta_{n}}{\alpha_{n}}=-1.559352647 \cdot 10^{-11} = \epsilon.
\end{eqnarray}
This small error can become a concrete issue when trying to obtain geostrophic balance for T-REXI using the SH method.
Our ad hoc solution was to rescale the $\beta_n$ coefficients with this residual $\epsilon^{-1}$.
Strongly improved results are obtained using the renormalized $\beta_n$ coefficients, with T-REXI errors of $O(10^{-12})$, as are shown in shown  Fig.\,\ref{fig:geostrophic_benchmark_with_normalization}.
We close this section with a brief discussion of T-REXI error growth in the geostrophic balance case.
Let $1-\epsilon$ be the REXI response to the geostrophic modes at $x=0$ which is expected to be $1$ and $\epsilon > 0$.
After $N$ time steps, the max.\,response error to the geostrophic modes is $\left( 1-\epsilon \right)^N$ and the T-REXI error growth can be estimated as
$
	\epsilon_{T} = \left(1-\epsilon\right)^N - 1
$.
The number of T-REXI time steps leading to an error of $\epsilon_T$ is then given by
$
	N = \log_{1+\epsilon}\left|1-\epsilon_T\right|
$.
Thus, a cumulative error $\epsilon_T = 10^{-6}$ equivalent to single precision would therefore be introduced after about $60000$ T-REXI time steps, an issue only relevant for simulations which demand highly accurate time integration of stationary modes.

\subsection{Propagation of Gaussian bumps}
\label{sec:propagation_gaussian_bumps}

In this section, we compare wave propagation in T-REXI to the n-th order Runge-Kutta  (RKn), and Crank-Nicolson (CN) time stepping methods.
The test case that we have created is designed to help analyze and visualize dispersion errors in numerical methods.
The initial state begins at rest ($u = v = 0$) with a displacement of the surface height with a Gaussian-shaped function, or ``bump'' of the form
\begin{eqnarray*}
	d(\lambda_1, \phi_1, \lambda_2, \phi_2)	&=&
				acos\left(
					\sin\phi_1 \sin\phi_2  +
					\cos\phi_1 \cos\phi_2 \cos(\lambda_1-\lambda_2)
				\right)\\
	\psi(\lambda, \phi, \lambda_c,\phi_c,p) &=& 
	\exp \left(
		-d(\lambda_c,\phi_c,\lambda,\phi)^2  \cdot p
	\right)
	0.1 \overline{H}
\end{eqnarray*}
with the superscript $c$ denoting the coordinates of the Gaussian bump and $p$ controls its width.
To cover a larger area of the power spectrum, we use a superposition of three Gaussian bumps of different widths:
\begin{eqnarray*}
	H(\lambda,\phi) &=& \overline{H} + \psi(\lambda, \phi, 0.2 \pi, \frac{1}{3}\pi, 20)+
	 	\psi(\lambda, \phi, 1.2 \pi, -\frac{1}{5}\pi, 80)+
		\psi(\lambda, \phi, 1.6 \pi, -\frac{1}{4}\pi, 360).
\end{eqnarray*}
The simulation is executed for 1.5 days using earth-like physical parameters, and a $T128$ SH truncation scheme was used (see Sec.\,\ref{sec:spherical_harmonics}).
During the test, the Gaussian bumps propagate over the sphere as disturbances, and finally reassembling themselves near their original positions.
For all studies we generate a reference height ($H$) solution using an explicit RK4 method and a time step size of $50s$.
In the following results, the reference solution is plotted as dashed isolines at intervals of $\Delta H = 30m$ relative to the average height $\overline{H}$.

\fboxsep=0pt
\fboxrule=2pt
\newcommand{\zoomedareaimage}[1]{\begin{tikzpicture}[remember picture,overlay]
    \node at (-7cm,1.9cm) {
			\fcolorbox{red}{red}{\includegraphics[width=3.5cm,trim={12.2cm 1.7cm 2.5cm 4.3cm},clip]{#1}}
    };
\end{tikzpicture}
}

\begin{figure}
	\begin{center}
	~~~~~\includegraphics[width=0.7\textwidth]{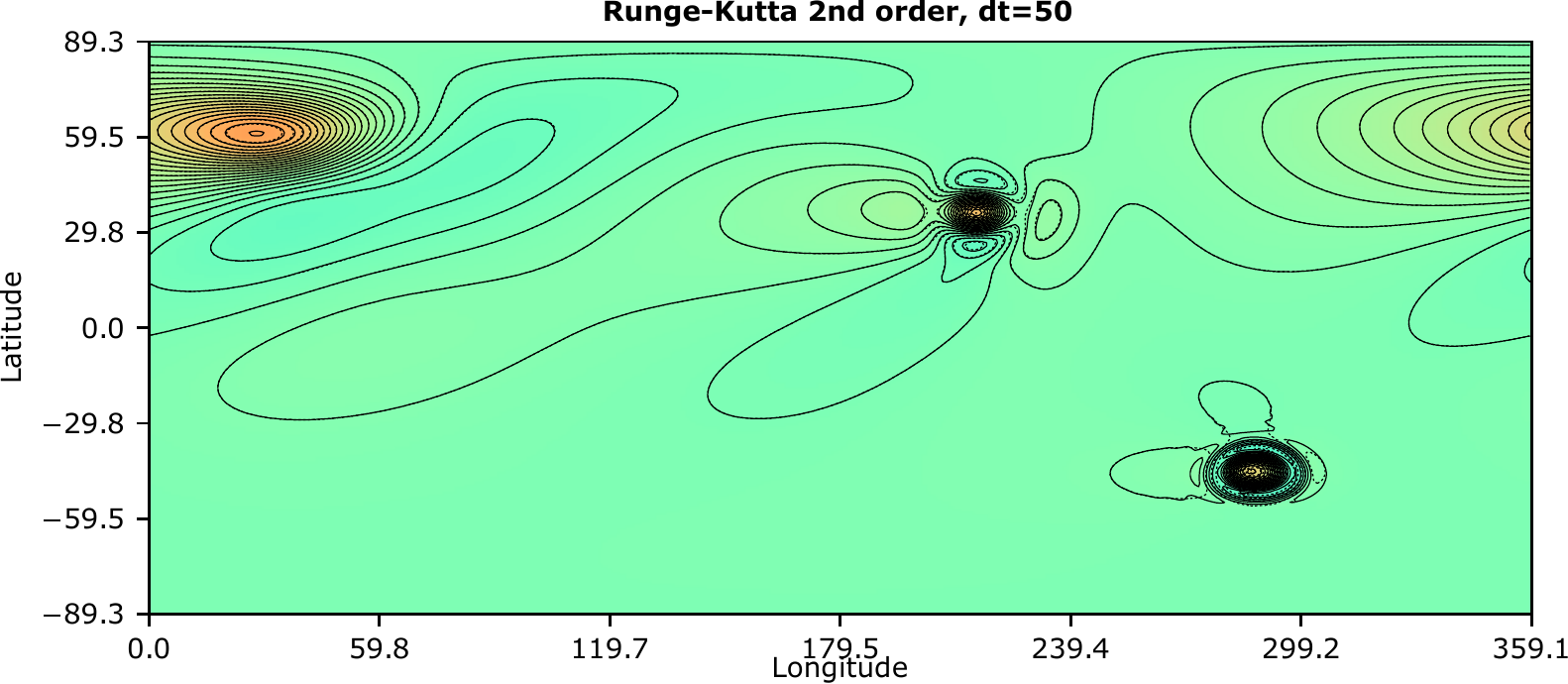}
	\zoomedareaimage{benchmarks_sphere_sph_rexi_linear_paper_gaussian_ts_comparison_earth_scale/result_script_g9_80616_h10000_f7_292e-05_a6371220_fsph0_u0_U0_tsm_l_erk_tso2_tsob1_C000050_REXITER_m00000256_h0_15_nrm1_new.pdf}
	~~\includegraphics[width=0.065\textwidth]{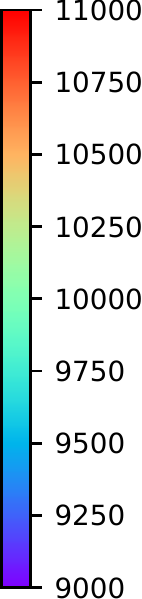}
	\newline

	\includegraphics[width=0.7\textwidth]{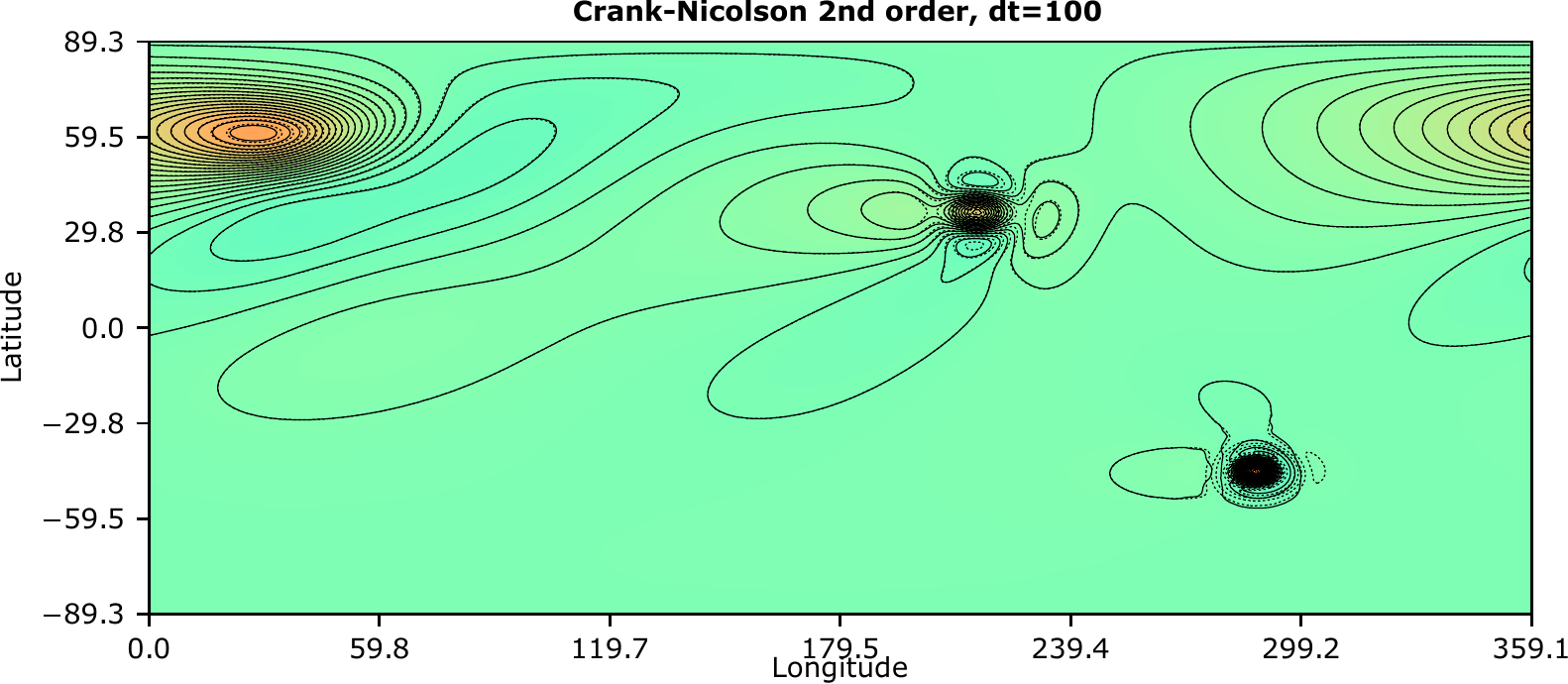}
	\zoomedareaimage{benchmarks_sphere_sph_rexi_linear_paper_gaussian_ts_comparison_earth_scale/result_script_g9_80616_h10000_f7_292e-05_a6371220_fsph0_u0_U0_tsm_l_cn_tso2_tsob1_C000100_REXITER_m00000256_h0_15_nrm1_new.pdf}
	\newline
	\end{center}

	\caption{
	\label{fig:gaussian_bump_std_ts_comparisons}
	Comparison of the surface height field produced for two different time stepping methods after $1.5$ days. The three 
	Gaussian bumps have closely reassembled themselves. The reference solution (dashed lines) is compared to the results of each 
	method (solid isolines).
	The red rectangle zooms in on one of the bumps closest to the south pole.
	Results with $\Delta t = 50s$ closely resemble the reference solution (only $2^{nd}$ order Runge-Kutta (RK2) is shown 		here).
	For the Crank-Nicolson (CN) method with a 100 second time step, dispersion errors (expressed by non-matching isolines) are clearly visible.
	Further increases of the time step size led to instabilities for the RK2 method, and even worse mismatches of the 
	isolines for CN.
	}
\end{figure}

In Fig.\,\ref{fig:gaussian_bump_std_ts_comparisons}, the dispersion errors for commonly-used, $2^{nd}$ order time stepping methods are compared to the 4th order (RK4) reference solution. For a time step size of $\Delta t = 50s$, all $2^{nd}$ order in time methods give reasonable solutions, although errors can be observed even for this relatively small time step size (only the RK2 results (top panel) are shown). 
Doubling the time step to $\Delta t=100s$ leads to visible instabilities\footnote{We would like to mention here that RK2 has known problems for oscillatory problems, however provided stable results within the chosen time integration interval.} for RK2, and CN methods show significantly increased dispersion errors, visible as non-matching isolines in the lower panel.

\begin{figure}
	\begin{center}
	~~~~~\includegraphics[width=0.7\textwidth]{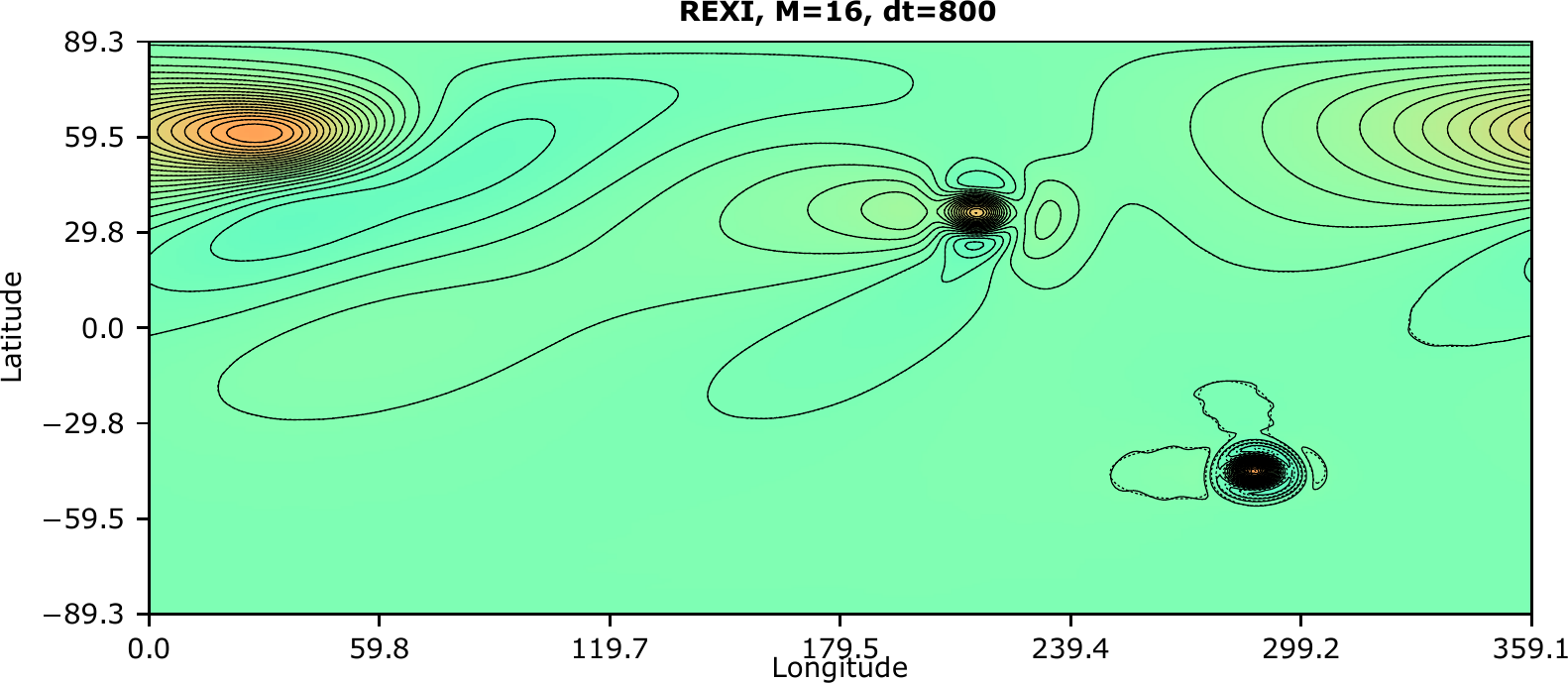}
	\zoomedareaimage{benchmarks_sphere_sph_rexi_linear_paper_gaussian_ts_comparison_earth_scale/result_script_g9_80616_h10000_f7_292e-05_a6371220_fsph0_u0_U0_tsm_l_rexi_tso0_tsob1_C000800_REXITER_m00000016_h0_15_nrm1_new.pdf}
	~~\includegraphics[width=0.065\textwidth]{benchmarks_sphere_sph_rexi_linear_paper_gaussian_ts_comparison_earth_scale/legend.pdf}\newline
	\newline
	\includegraphics[width=0.7\textwidth]{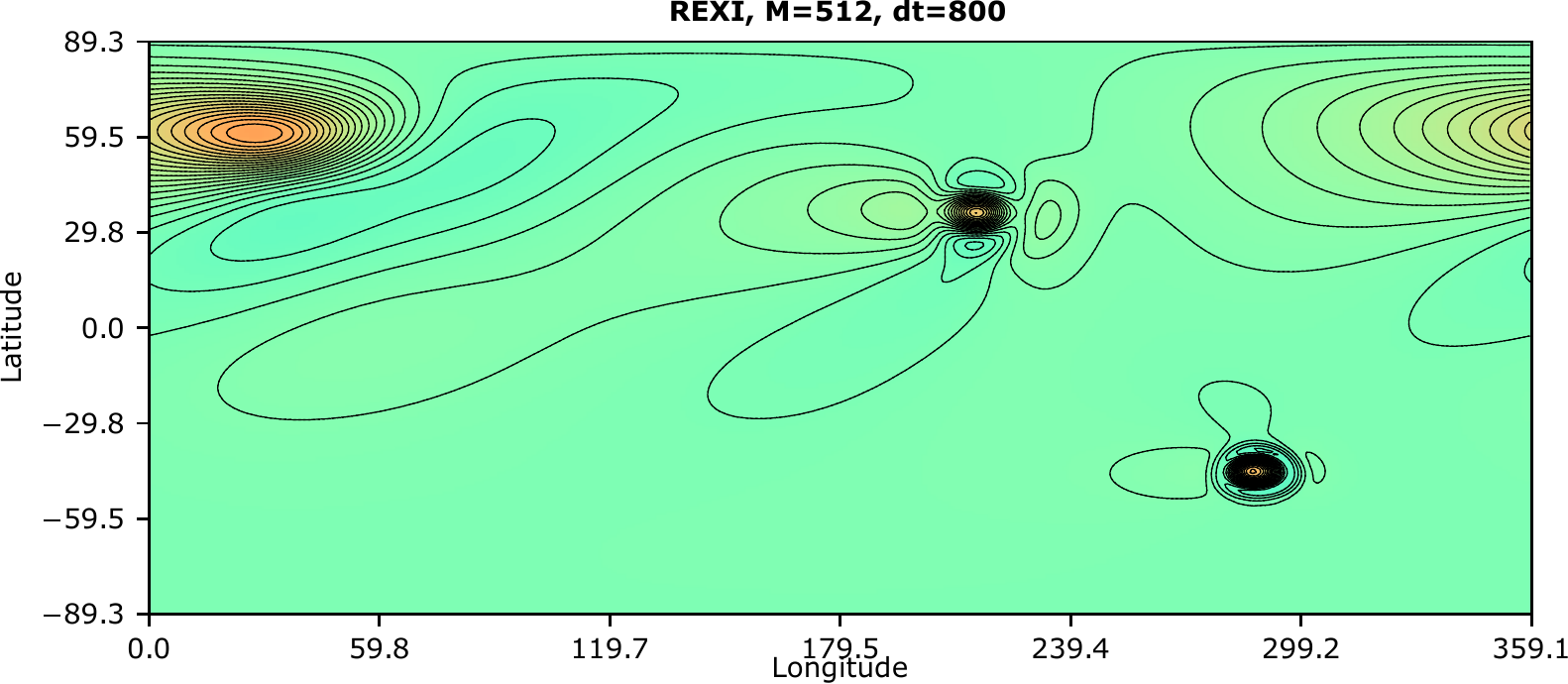}
	\zoomedareaimage{benchmarks_sphere_sph_rexi_linear_paper_gaussian_ts_comparison_earth_scale/result_script_g9_80616_h10000_f7_292e-05_a6371220_fsph0_u0_U0_tsm_l_rexi_tso0_tsob1_C000800_REXITER_m00000512_h0_15_nrm1_new.pdf}
	\newline
	\end{center}

	\caption{
	\label{fig:gaussian_bump_rexi_comparisons_dt800}
	The surface height field produced after $1.5$ simulated days using T-REXI time integration scheme is compared to the RK4 reference solution, again 		printed with dashed isolines. A time step of $800s$ was used for T-REXI.
	The first plot shows considerable errors in the propagation due to an insufficient number of T-REXI poles (M=16 in this case).
	The second plot shows very accurate results for T-REXI with $M = 512$, compared to RK2 and CN results from Fig.\,\ref{fig:gaussian_bump_std_ts_comparisons}.
	}
\end{figure}

\begin{figure}
	\begin{center}
	~~~~~\includegraphics[width=0.7\textwidth]{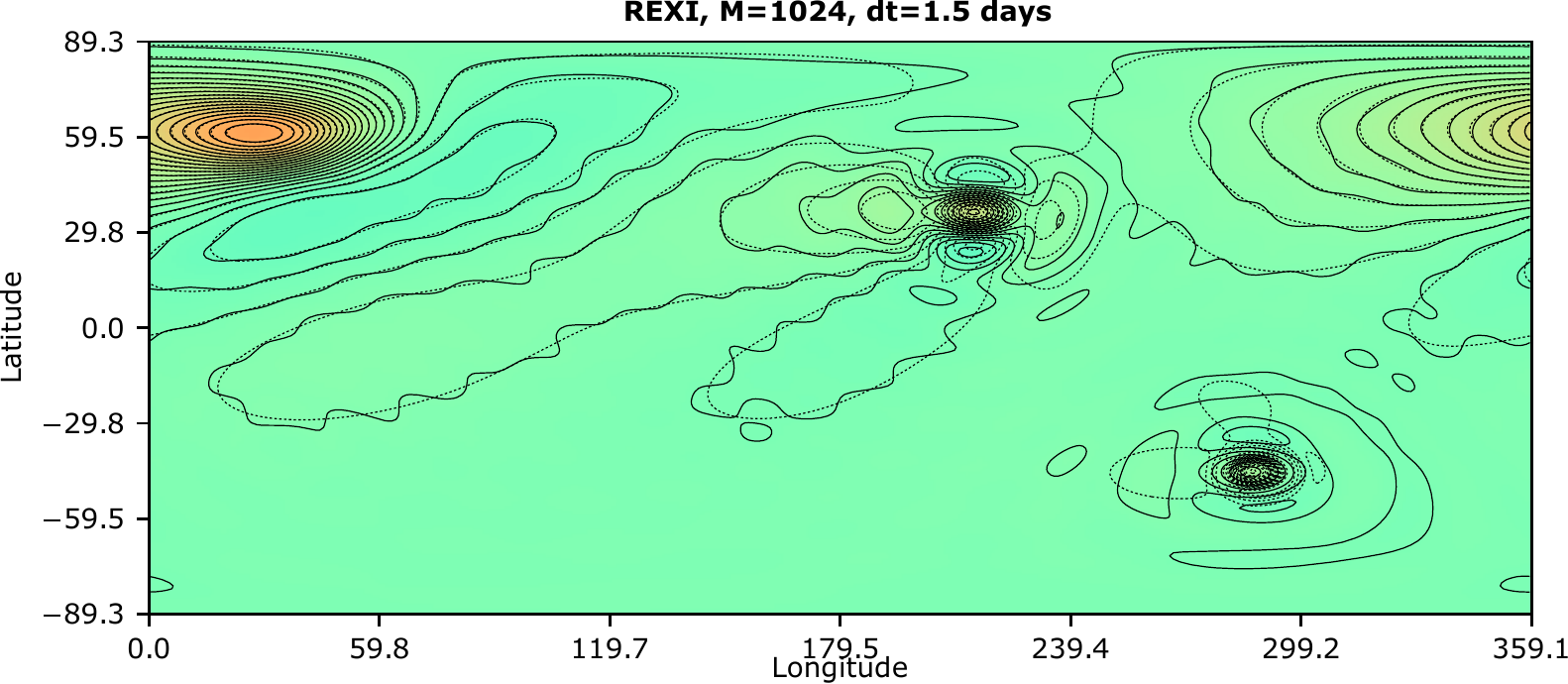}
\zoomedareaimage{benchmarks_sphere_sph_rexi_linear_paper_gaussian_ts_comparison_earth_scale/result_script_g9_80616_h10000_f7_292e-05_a6371220_fsph0_u0_U0_tsm_l_rexi_tso0_tsob1_C129600_REXITER_m00001024_h0_15_nrm1_new.pdf}
	~~\includegraphics[width=0.065\textwidth]{benchmarks_sphere_sph_rexi_linear_paper_gaussian_ts_comparison_earth_scale/legend.pdf}\newline
	\newline
	\includegraphics[width=0.7\textwidth]{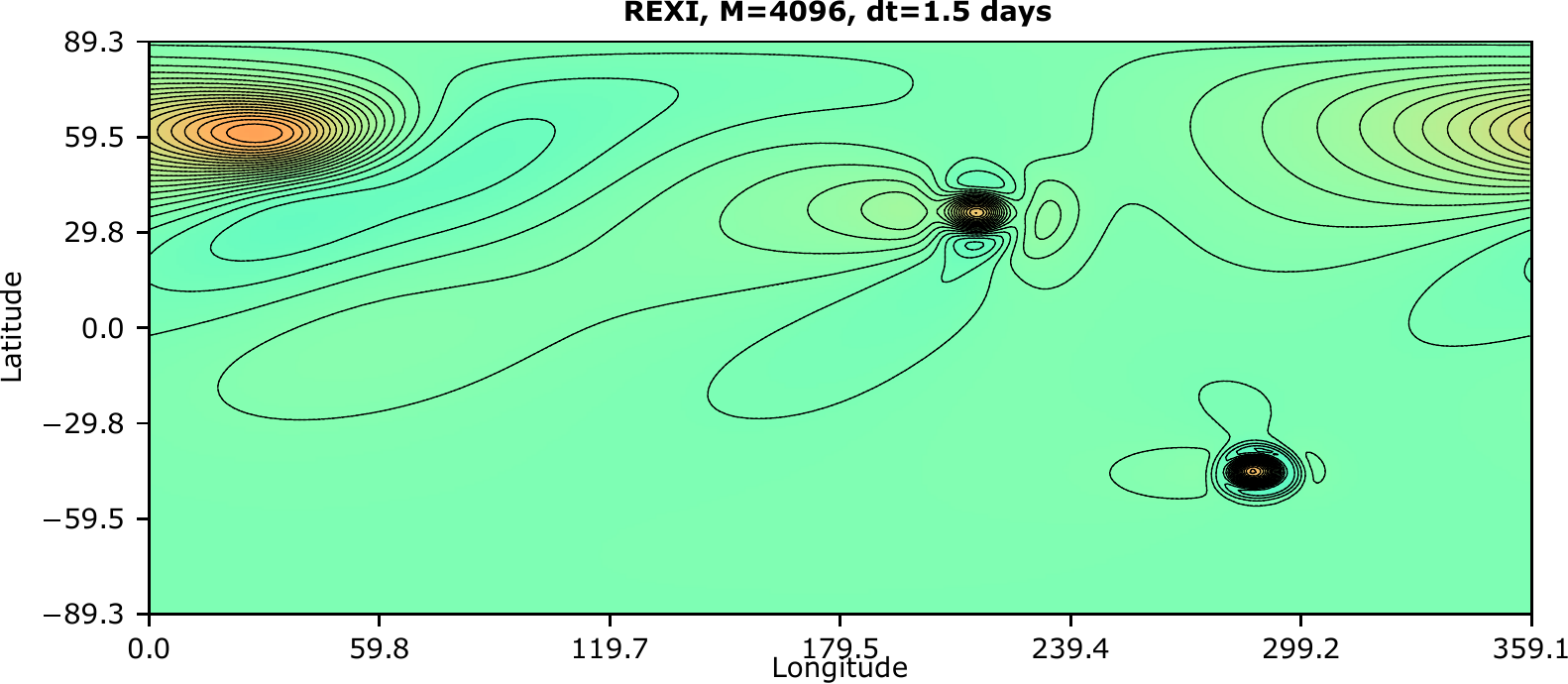}
	\zoomedareaimage{benchmarks_sphere_sph_rexi_linear_paper_gaussian_ts_comparison_earth_scale/result_script_g9_80616_h10000_f7_292e-05_a6371220_fsph0_u0_U0_tsm_l_rexi_tso0_tsob1_C129600_REXITER_m00004192_h0_15_nrm1_new.pdf}
	\newline
	\end{center}

	\caption{
		\label{fig:gaussian_bump_rexi_comparisons_dt129600}
		Same as Fig.\,\ref{fig:gaussian_bump_rexi_comparisons_dt800}, but with T-REXI time step set to $1.5$ days.
		The RK4 reference solution is computed and plotted as before.
		The first plot (upper panel) shows significant errors in the solution with the T-REXI $M=1024$: accuracy is restored in the lower panel
		using $M=4192$.The results obtained are of higher accuracy than the solution computed with the RK2 method.
	}
\end{figure}

We now turn to comparisons of the T-REXI method with the RK4 reference solution. Fig.\,\ref{fig:gaussian_bump_rexi_comparisons_dt800} shows T-REXI solutions $\Delta t = 800s$, already 16 times larger than RK2.
The T-REXI solution in the upper panel ($M=16$) shows significant errors. This is due to using a small number of poles, insufficient to approximate the eigenvalue spectrum.
However, increasing the number of T-REXI integration poles to $M=512$ leads to a highly accurate solution (lower panel).

In Fig.\,\ref{fig:gaussian_bump_rexi_comparisons_dt129600}, we compare T-REXI with the RK4 reference solution when T-REXI takes one enormous 1.5 day time step.
For the T-REXI $M=1024$ case, significant errors are again generated (top panel).
However, increasing the number of T-REXI integration poles to $M=4098$ again restores agreement with the reference solution.
Perhaps remarkably, T-REXI is able to yield a higher accuracy solution than the RK2 method (top panel of Fig.\,\ref{fig:gaussian_bump_std_ts_comparisons}), using a $2592$ times larger time step.

\subsection{Numerical dispersion analysis}
\label{sec:eigenvalue_analysis}

In this section we provide an in-depth numerical analysis of the dispersion relations of T-REXI vs.\,other time stepping methods.
Explicit and implicit methods are known to suffer from either accelerating or decelerating the dispersion speeds of waves (see \cite{hoskins1975multi,durran2010numerical,lynch2006emergence}).
In this section we use a mode analysis to reveal dispersion properties and errors, see e.g.\,\cite{thuburn2009numerical,weller2012computational,peixoto2016accuracy}.
Dispersion errors due to spatial discretization are avoided by our choice of the global SH method, which allows us to cleanly isolate the dispersion errors arising from the time stepping methods alone.
Due to computational complexity of an Eigenvalue decomposition, this study used a reduced T16 wave number truncation scheme (see Section \ref{sec:spherical_harmonics}).

We analyze the dispersion relations over a sufficiently large time such that dispersion errors in the time stepping method itself are revealed.
Given the matrix $E$ which integrates $\textbf{U}_n$ to $\textbf{U}_{n+1}$, we obtain the dispersion relations in $\Lambda$ by using an eigenvector/value formulation for exponential integrators with
$
	\textbf{U}_{n+1}=E \textbf{U}_{n} = Q e^{\Delta t\Lambda}Q^{-1}\textbf{U}_{n}.
$
Taking the logarithm of the eigenvalues of $E$ then reveals details on the dispersion modes.
Special attention is required since the logarithm on complex numbers is not bijective: the eigenvalues obtained might not be related to the real ones but to ones shifted by multiples of $2 \pi$, yet still yielding the correct results.
We used a maximum time integration interval of $\Delta t = 400s$ which assures a bijective property.
The linear operator matrix $E$ itself is obtained by iterating over all modes of the state vector $\textbf{U}_n$.
In each iteration, only the current mode is activated and a time step is executed.
The resulting state vector $\textbf{U}_{n+1}$ then represents one column of the linear operator matrix $E$.

\begin{figure}
	\begin{center}
	\includegraphics[width=0.8\textwidth]{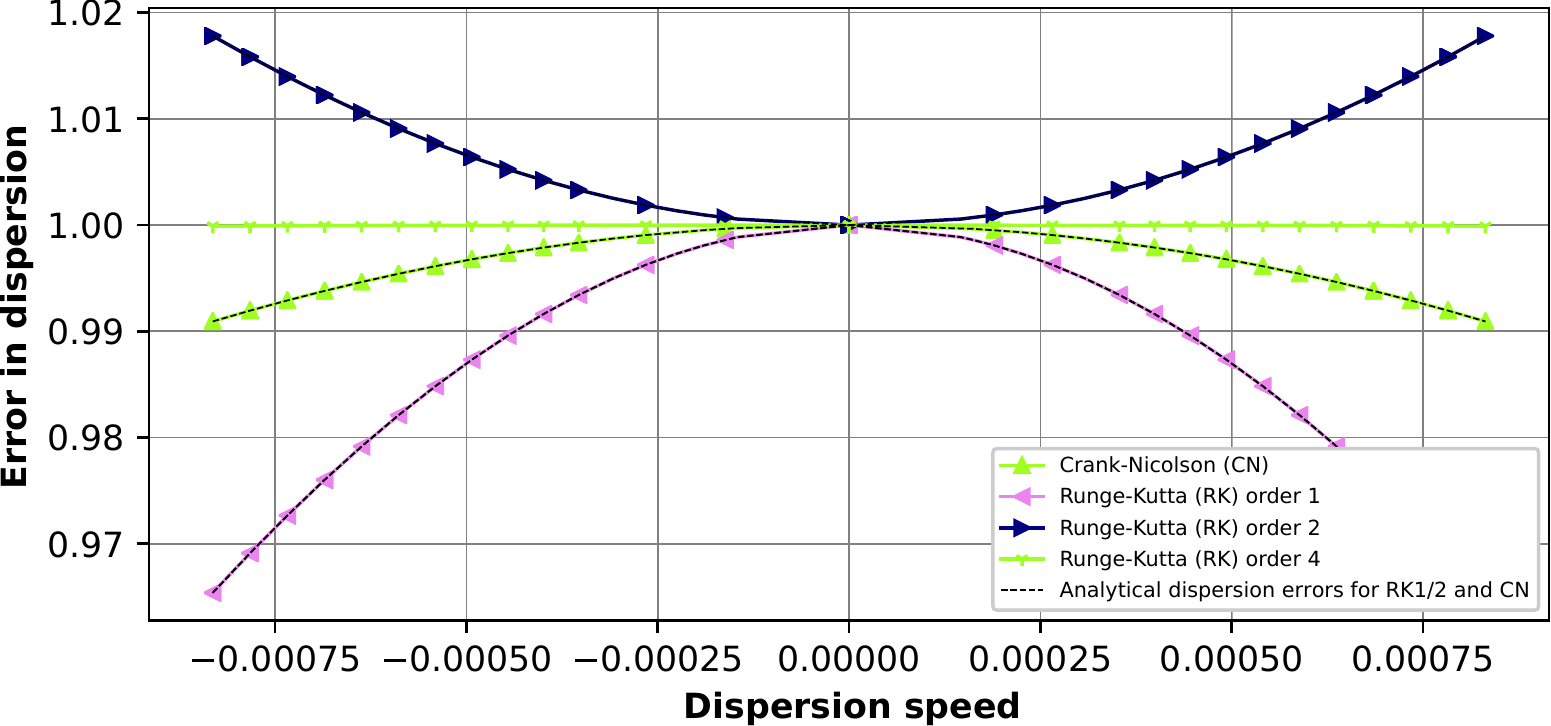}
	\end{center}
	\caption{	
\label{fig:a_output_combined_fsphere1_rexihalf1_rexinorm0_rexiextmodes02_STDTS}
		Relative phase errors for Runge-Kutta of order 1, 2, and 4 and Crank-Nicolson based on the linear SWE computed with SH. We can observe significant errors for $2^{nd}$ order accurate time stepping methods for fast moving wave waves (see also \cite{durran2010numerical}).
		Markers are set for every $10^{th}$ mode.
	}
\end{figure}
We performed studies on the f-sphere with the results provided in Fig.\,\ref{fig:a_output_combined_fsphere1_rexihalf1_rexinorm0_rexiextmodes02_STDTS}.
In general, there are significant errors for $1^{st}$ and $2^{nd}$ order accurate methods.
The RK2 method accelerates waves whereas all other considered methods decelerate them (see also \cite{durran2010numerical}).
The $4^{th}$ order accurate Runge-Kutta method results in relatively small errors which are significantly smaller than those for $2^{nd}$ order accurate methods.

\begin{figure}
	\begin{center}
	\includegraphics[width=0.8\textwidth]{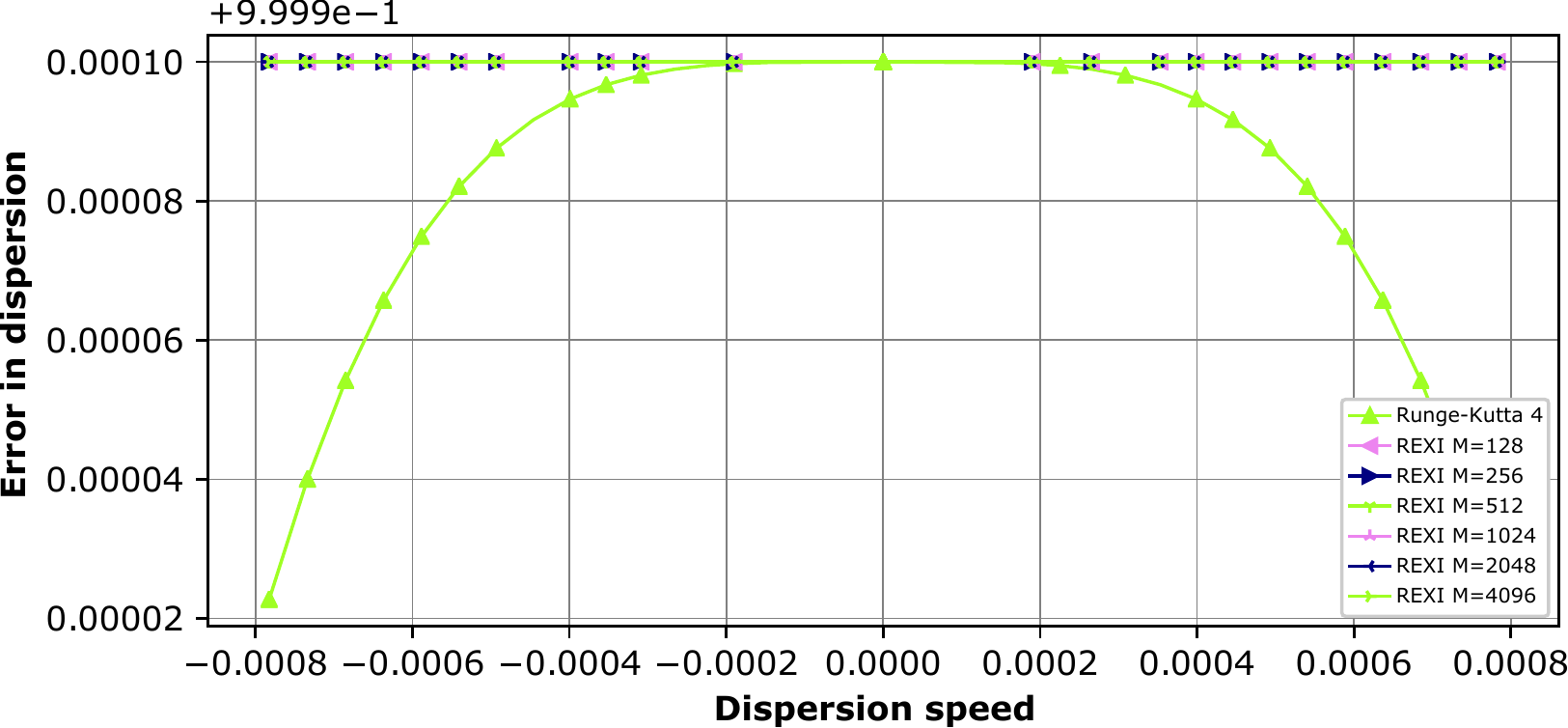}
	\end{center}
	\caption{	\label{fig:a_output_combined_fsphere1_rexihalf1_rexinorm0_rexiextmodes02_REXI}
		Relative phase errors for T-REXI and RK4 with the linear SWE computed with SH.
		We can observe that
        T-REXI is able to cope significantly better with fast waves.
		Markers are set for every $10^{th}$ mode.
	}
\end{figure}
To show the potential of T-REXI, we compare it to RK4 in Fig.\,\ref{fig:a_output_combined_fsphere1_rexihalf1_rexinorm0_rexiextmodes02_REXI}.
Here, RK4 still results in larger errors for fast moving waves.
In contrast, T-REXI does not show increased wave dispersion errors for faster moving waves in the given example.

\subsection{Performance comparison with massively parallel REXI}
\label{sec:performance_comparison}

Next, we compare the computational performance of T-REXI with conventional time stepping methods.
All performance results are conducted on the Cheyenne supercomputer\cite{cheyenne}.
The performance metric we adopt for our performance intercomparison is based on the wall clock time required to perform a full time integration of $1.5$ days.
The benchmark setup is identical to the propagating Gaussian bumps from Section \ref{sec:propagation_gaussian_bumps}.
The reference solution was computed with a Runge-Kutta 4 method and $\Delta t = 50s$ and the errors are computed with the $\ell_{\infty}$ norm to the reference height.
We used RK2 and CN as conventional time stepping methods and executed them using only a single core on an exclusively reserved NUMA domain.
Using more cores would be only beneficial for a significantly increased workload (e.g.\,by significantly increasing the resolution or extending it to the vertical).
For RK2, time step sizes $\Delta t \geq 100s$ turned out to be unstable and, despite being stable, for CN time step sizes of $\Delta t \geq 150s$ resulted in significant errors, see Sec.\,\ref{sec:propagation_gaussian_bumps}.
For T-REXI, we conducted different studies ($\Delta t=\{800s, 1600s, 1.5 days\}$, $M=\{512, 1024, 4096\}$) and distributed the PDEs of each REXI term equally across $N=\{2^i | i \leq 0 < 10\}$ compute ranks.
We use only one rank per socket (two per compute node) to maximize the available bandwidth to solve for each term in the T-REXI formulation.
Hence, the maximum number of used compute nodes on Cheyenne is $\frac{2^9 \text{~MPI ranks on one socket}}{2 \text{~sockets per node}}=256$.

\begin{figure}
	\begin{center}
		\includegraphics[width=0.8\textwidth]{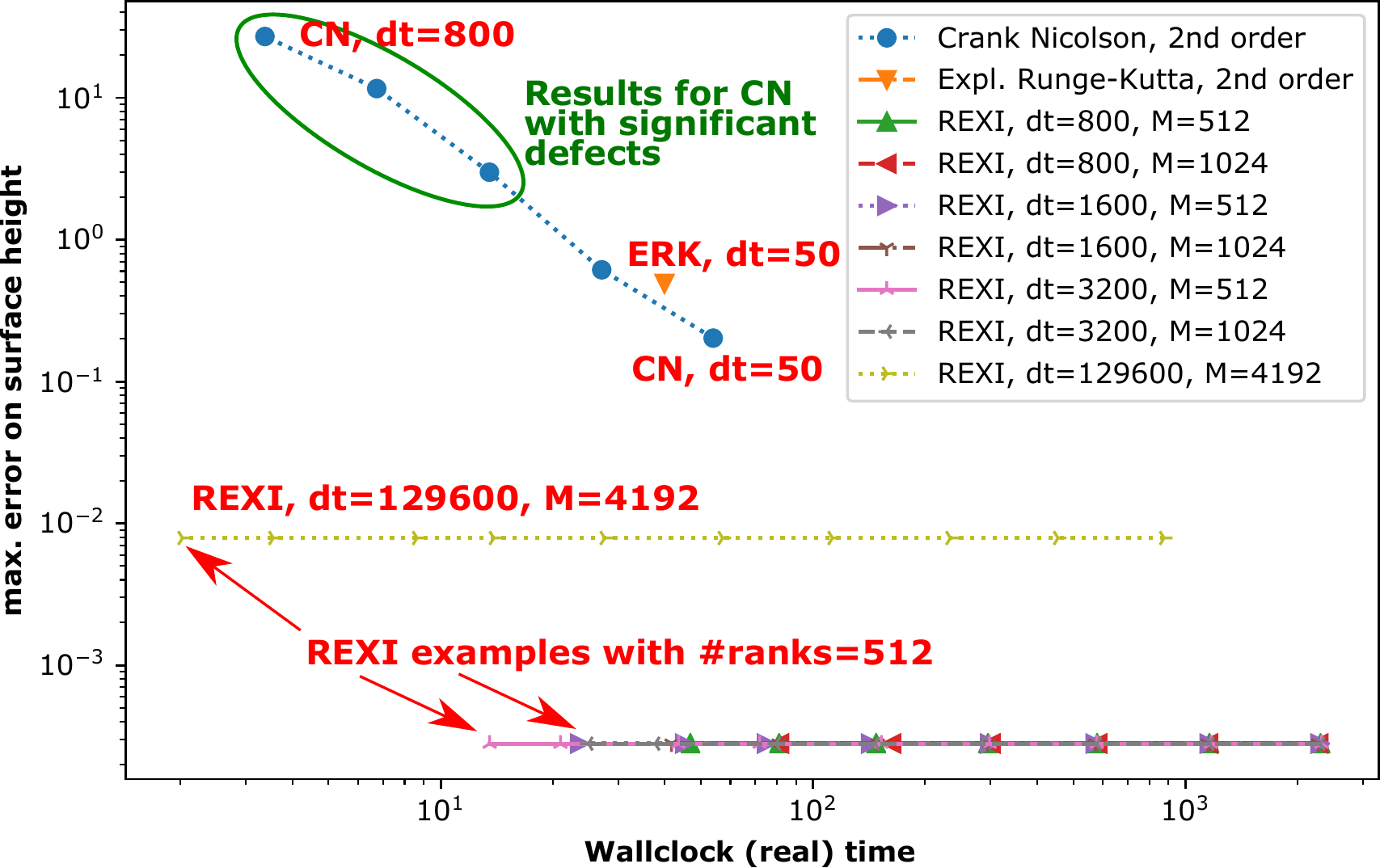}
	\end{center}
	\caption{
	\label{fig:performance_comparison_cheyenne}
		Wallclock vs.\,error on the difference in surface height comparing explicit RK, Crank-Nicolson and T-REXI time stepping methods.
		We can observe that runs based on T-REXI always result in improved accuracy compared to the other time stepping methods.
		Based on the results in a previous section, it is important to mention that Crank-Nicolson time stepping results with $dt > 100s$ are of no use due to significant errors.
		T-REXI time stepping uses up to 512 additional compute ranks and shows significant improvements regarding the time-to-error compared to other time stepping methods.
	}
\end{figure}

The results are given in Fig.\,\ref{fig:performance_comparison_cheyenne}.
Using T-REXI always results in errors which are lower than for the conventional time stepping methods.
Therefore, for the studied conventional methods only a further reduction of time step size would make them competitive, however also increasing their wallclock time.
Given the best wallclock time of $27.1s$ for the conventional TS method CN with $\Delta t = 100s$ and comparing it to the best REXI method with $M=512$ and $\Delta t = 3200$ with a wallclock time of $13.5s$ reveals a speedup of $2.0\times$.
Even if we don't expect that time step sizes of $1.5days$ with a wallclocktime of $2.02s$ can be used
once including the non-linearities, we would like to mention the potential wallclock speedup of $13.4\times$ if a way to incorporate the non-linearities can be found as part of future research.
We also like to mention that the scalability limitation with $M=4096$ T-REXI terms was not yet reached since we only used $512$ compute ranks.

\section{Summary and future work}

The apparent advantages of the Terry Haut et al.\,\cite{Haut2015} rational approximation of exponential integrators (T-REXI) method are that it allows, for linear oscillatory operators, (a) arbitrarily long time steps, and (b) a realization with a sum over the solutions to a series of embarrassingly parallel solvers of the form
$
	\textbf{U}_{n+1} \approx \sum_i \beta_i (\alpha_i + \Delta t L)^{-1} \textbf{U}_n.
$ 
In this work we applied T-REXI to the linear terms of the shallow-water equations (SWE) on the rotating sphere.
Applying T-REXI to all linear terms of this SWE requires coping with the Coriolis term induced by the rotating sphere and leads to additional challenges.
These have been overcome through a formulation which solves for the geopotential alone, reducing the complexity of the system of equations by a factor of three.
Once the geopotential is obtained, it is straightforward to directly compute the velocities.
Using Spherical Harmonic (SH) method enables the formulation of T-REXI's $\left( \alpha_i + \Delta t L \right)^{-1}$ term for the rotating SWE as a low-bandwidth matrix in spectral space with the bandwidth independent of the resolution: this leads immediately to the required efficient direct solver.
Additionally, we can avoid dispersion errors (up to numerical precision) of gravity as well as Rossby waves which are both of significant importance in the atmosphere.

Three test cases relevant to atmospheric simulations were conducted for this approach: geostrophic balance, Gaussian breaking dams, and wave dispersion. 
The geostrophic balance stability test case revealed small SH modal errors which a simple normalization of T-REXI coefficients resolved.
We assessed T-REXI's numerical performance with a propagation of three differently-sized Gaussian bumps across the earth over 1.5 simulated days, and compared the dispersion errors of the T-REXI scheme to two widely-used time stepping methods (Runge-Kutta 2, Crank-Nicolson).
T-REXI was able to take a single 1.5 day time step which led to smaller solution errors than those achieved by the established methods requiring much smaller time steps.
The price to pay for this was to increase the number of T-REXI terms to $4096$, however exposing the potential to parallelise over these terms.

The ability of a numerical method to accurately reproduce the dispersion of waves in the SWE is of particular interest for climate and weather simulations, where tracking atmospheric effects and their interactions accurately over long time integration intervals is required.
Here, wave dispersion relations for widely-used time stepping methods were extracted and particular defects for fast moving waves observed.
T-REXI avoided significant wave dispersion errors while taking larger time step sizes.
All our studies demonstrated T-REXI's superior properties regarding its dispersion relations.

We conducted performance studies on Cheyenne supercomputer by parallelizing over the T-REXI terms.
Using a time step size which is $32$ time larger than using a Crank-Nicolson method, the wallclock time is reduced by $2.0 \times$ and also the errors are significantly reduced.
Taking T-REXI to its extreme, we also performed one large $1.5 day$ time step with $M=4096$ using T-REXI which resulted in a significant reduction of the errors and yielded a speedup of $13.4\times$, with the scalability limitation not yet reached.

In summary, our results show that T-REXI can be successfully extended to the rotating sphere, can take very large time steps in the case of the stiff, linear oscillatory terms of the SWE, and is competitive to other time stepping schemes.
The embarrassingly parallel set of matrix inversion problems at the heart of T-REXI is well aligned with modern computing technology trends.
However, some limitations and obvious extensions of the current work are worth noting.
First, methods based on SH have well recognized numerical and computational limitations that have led many in the atmospheric community to turn to other approaches.
Therefore, the application of T-REXI to other numerical schemes using scalable iterative solvers should be investigated.
Second, while our focus on applying T-REXI to the linear parts of the SWE on the rotating sphere is an appropriate first step, we fully expect that extensions of this work to the fully non-linear SWE are required, and would likely lead to time step restrictions on the T-REXI scheme.
Such extensions to the non-linearities can be accomplished in different ways such as Strang-splitting, non-linear exponential integrators and other parallel-in-time methods (ParaEXP\cite{Gander2013}, aPinT\cite{Haut_Wingate_14}).
Unfortunately, the severity of these restrictions regarding wallclock time-to-solution including the non-linear interactions and how to overcome them for the SWE on the rotating sphere is not yet researched.
These challenges remain, therefore, important topics for future research in the development of (REXI-based) parallel-in-time methods.

\acks

We'd like to acknowledge computation time for early computational studies on the Yellowstone cluster \cite{yellowstone2012} and the MAC cluster at TUM.
We would like to acknowledge high-performance computing support from Cheyenne \cite{cheyenne} (doi:10.5065/D6RX99HX) provided by NCAR's Computational and Information Systems Laboratory, sponsored by the National Science Foundation.
Martin Schreiber likes to thank
Pedro S.\,Peixoto,
Nathanaël Schaeffer,
Nils Wedi,
Terry Haut,
Jemma Shipton,
Houjun Wang,
and Beth Wingate
for various discussions over the last 2.5 years.
We'd like to thank the anonymous reviewers for their valuable feedback which strongly improved the focus of this paper on the T-REXI method.

\bibliographystyle{wileyj} 
\bibliography{bibliography}

\end{document}